\documentstyle{article}
\begin{document}
\hoffset -1.7cm
\pagestyle{headings}

\setlength{\hsize}{14.5cm}
\setlength{\vsize}{20cm}
\renewcommand{\baselinestretch}{1.2}

\title{\bf Analysis of $SU(3)_c\otimes SU(3)_L\otimes U(1)_X$ local gauge
theory}
\author{{\bf William A. Ponce}\\
{\it Instituto de F\'\i sica, Universidad de Antioquia,}\\
{\it A.A. 1226, Medell\'\i n, Colombia.}
\and
{\bf Ju\'an B. Fl\'orez}\\
{\it Depto. de  F\'\i sica, Universidad de Nari\~no}\\
{\it Pasto, Colombia.}
\and
{\bf Luis A. S\'anchez}\\
{\it Escuela de F\'\i sica, Universidad Nacional de Colombia}\\
{\it A.A. 3840, Medell\'\i n, Colombia.}}
\date{}
\maketitle

\begin{abstract}

Six different models, straightforward extensions of the standard model to
$SU(3)_c\otimes SU(3)_L\otimes U(1)_X$, which do not contain particles
with exotic electric charges are presented. Two of the models are one
family and four are three family models. In two of the three family models
one of the families transforms different from the others, and in the other
two all the three families are different.
\end{abstract}

\pagebreak
\large

\section{Introduction}

The remarkable experimental success of the standard model (SM) local gauge
group $G_{SM}\equiv SU(3)_c\otimes SU(2)_L\otimes U(1)_Y$ with the flavor
sector $SU(2)_L\otimes U(1)_Y$ hidden\cite{sm} and $SU(3)_c$
confined\cite{su3}, lies in its accurate predictions at energies below a
few hundreds GeV. However, the SM is not the only model for which this is
true and many physicists believe that it does not represent the final
theory, but serves merely as an effective theory originating from a more
fundamental one. So, extensions of the SM are always welcome.

One can extend the SM either by adding new fermion fields (adding a
right-handed neutrino field constitute its simplest extension), by
augmenting the scalar sector to more than one higgs representation, or
by enlarging the local gauge group. In this last direction,
$SU(3)_L\otimes U(1)_X$ as a flavor group has been studied previously by
many authors\cite{cha} who have explored many possible fermion and
higgs-boson representation assignments, either as identical replicas of
one family structures as in the SM \cite{fallet} or as a
multi-family structure\cite{fp,long} which points to a natural explanation
of the total number of families in nature.

With regard to the different models in Ref.\cite{fallet}, most of them are
plagued with physical inconsistencies such as gauge anomalies,
right-handed currents at low energies, unwanted flavor changing neutral
currents, violation of universality, etc.. The model in Refs.\cite{fp} for
three families of quarks and leptons is consistent with the low energy
phenomenology and it is anomaly free thanks to the introduction of quarks
with exotic electric charges $-4/3$ and $5/3$. On the other hand, the
model in Refs.\cite{long}, also for three families, is consistent with low
energy phenomenology and does not include particles with exotic electric
charges.

In this paper we present an analysis of the local gauge group
$SU(3)_c\otimes SU(3)_L\otimes U(1)_X$. We find six models which are
anomaly free, do not include fermions (quarks and leptons) with exotic
electric charges and are consistent with the low energy phenomenology.
Two of the models are one family models and are natural extensions of the
SM (one of them is an $E_6$ subgroup), while the other four are models for 
three families of quarks and
leptons; two of them, up to our knowledge, new in the literature.  The
models under consideration get their symmetries broken via the most
economical set of higgs fields. We analyze also the limit in which the
neutral currents reproduce the SM phenomenology.

Our paper is organized in the following way: In section two we introduce
the characteristics of the gauge group and present the six different
models mentioned above; in section three we describe the scalar sector
needed to break the symmetry; in section four we analyze the gauge boson
sector paying special attention to the two neutral currents and their
mixing, in section five we analyze the fermion masses for one particular
model and in the last section we give our conclusions.

\section{The model}

In what follows we assume that the electroweak gauge group is
$SU(3)_L\otimes U(1)_X\supset SU(2)_L \otimes U(1)_Y$. We also assume that
the left handed quarks (color triplets) and left-handed leptons (color
singlets) transform under the two fundamental representations of $SU(3)_L$
(the 3 and $3^*$). Two classes of models will be discussed: one family
models where the anomalies cancel in each family as in the SM, and family
models where the anomalies cancel by an interplay between the families. As
in the SM, $SU(3)_c$ is vectorlike.

All the models analyzed have the same gauge boson sector, but they differ
in their fermion content and may differ in the scalar sector too.

\subsection{One family models}
The most general expression for the electric charge generator in
$SU(3)_L\otimes U(1)_X$ is a linear combination of the three diagonal
generators of the gauge group
\begin{equation}\label{ch}
Q=aT_{3L}+\frac{2}{\sqrt{3}}bT_{8L}+XI_3,
\end{equation}
where $T_{iL}=\lambda_{iL}/2; \;\lambda_{iL}$ being the Gell-Mann matrices
for $SU(3)_L$ normalized as {\bf Tr}$(\lambda_i\lambda_j)=2\delta_{ij}$,
$I_3=Dg(1,1,1)$ is the diagonal $3\times 3$ unit matrix,
and $a$ and $b$ are arbitrary parameters to be calculated ahead. Notice
that we have absorbed an eventual coefficient for $X$ in its definition.

Now, having in mind the canonical iso-doublets for $SU(2)_L$ in one
family, we start by defining two $SU(3)_L$ triplets \\

\[\chi_L=\left(\begin{array}{c}u\\d\\q \end{array}\right)_L , \hspace{1cm}
\psi_L=\left(\begin{array}{c}e^-\\ \nu_e\\l\end{array}\right)_L, \]

\noindent where $q_L$ and $l_L$ are $SU(2)_L$ singlet exotic quark and lepton fields
respectively of electric charges to be fixed ahead. This structure implies
that $a=1$ in Eq.(\ref{ch}) and one gets a one-parameter set of models.
Now if the \{$SU(3)_L,\; U(1)_X$\} quantum numbers for $\chi_L$ and
$\psi_L$ are \{$3,X_\chi$\} and \{$3^*,X_\psi$\} respectively, then
by using Eq.(\ref{ch}) we have the relationship:
\begin{equation}\label{chql}
X_{\chi}+X_\psi=Q_q+Q_l=-1/3,
\end{equation}
where $Q_q$ and $Q_l$ are the
electric charge values of the $SU(2)_L$ singlets $q$ and $l$ respectively,
in units of the absolute value of the electron electric charge.

Now in order to cancel the $[SU(3)_L]^3$ anomaly, two more $SU(3)_L$
lepton anti-triplets with quantum numbers \{$3^*,X_i\},\, i=1,2$, must
be introduced (together with their corresponding right-handed charged
components). Each one of those multiplets must include one $SU(2)_L$
doublet and one singlet of new leptons.
The quarks fields $u^c_L,\; d^c_L$ and $q^c_L$ color anti-triplets and
$SU(3)_L$ singlets, with $U(1)_X$ quantum numbers $X_u,\; X_d$ and $X_q$
respectively, must also be introduced in order to cancel the $[SU(3)_c]^3$
anomaly. Then the hypercharges $X_\alpha$ with $\alpha=\chi ,\psi
,1,2,u,d,q,...$ are fixed using Eqs. (\ref{ch}), (\ref{chql}) and the
anomaly constraint equations coming from the vertices
$[SU(3)_c]^2U(1)_X,\;[SU(3)_L]^2U(1)_X, \; [grav]^2U(1)_X$ and
$[U(1)_X]^3$, which are:
\begin{eqnarray} \nonumber
[SU(3)_c]^2U(1)_X &:& 3X_\chi+X_u+X_d+X_q=0\\ \nonumber
[SU(3)_L]^2U(1)_X &:& 3X_\chi+X_\psi+X_1+X_2=0\\ \nonumber
[grav]^2U(1)_X  &:& 9X_\chi+3X_u+3X_d+3X_q+3X_\psi+3X_1+3X_2+
                  \sum_{singl}X_{ls}=0\\  \nonumber
[U(1)_X]^3      &:&
9X_\chi^3+3X_u^3+3X_d^3+3X_q^3+3X_\psi^3+3X_1^3+3X_2^3+\sum_{singl}
X^3_{ls}=0, \end{eqnarray}
where $X_{ls}$ are the hypercharges of the right-handed charged lepton
singlets needed in order to have a consistent field theory.

What we have so far is an infinite number of possible models each one
characterized by the parameter $b$ in Eq.(\ref{ch}); the value of $b$ is
the key factor in determining the electric charge of the extra particles
in the several models to be presented. We are going to drastically limit
this number of possible models by imposing the constraint of excluding
models with particles with exotic electric charges; that is, we are going
to allow only models with quarks of electric charges $\pm 2/3$ and $\pm
1/3$ and leptons of electric charges $\pm 1$ and 0. We will see that this
requirement render us with only two different sets of models $(b=\pm
1/2)$ with equivalent gauge sectors.

\subsubsection{Model A}
Let us start with a model with an extra down type quark $q=D$ of electric
charge $Q_q=Q_D=-1/3$ ($b=1/2$) which in turn implies $Q_l=0$, that is,
$l_L$ is a new neutral lepton $N^0_{1L}$.  Eq.(\ref{ch}) then implies
$X_q=X_d=1/3$, $X_u=-2/3$, which combined with the anomaly constraint
equations and Eq.(\ref{chql}) gives $X_\chi=0,\;  X_\psi=-1/3,\;
\sum_{singl}X_{ls} =0$ and $X_1+X_2=1/3$. By demanding for leptons of
electric charges $\pm 1$ and 0 only, we have for the simplest solution
that $X_1=-1/3,\;  X_2=2/3$ and $X_{ls}=0$, with this last constraint
meaning that we do not need right-handed charged leptons in our simplest
anomaly-free model.

Putting all this together we end up with the following multiplet
structure for this model:
\[\begin{array}{||c|c|c|c||}\hline\hline
\chi_L=\left(\begin{array}{c}u\\d\\D \end{array}\right)_L & u^c_L & d^c_L&
D^c_L \\ \hline
(3,3,0) & (3^*,1,-{2\over 3}) & (3^*,1,{1\over 3}) &
(3^*,1,{1\over 3}) \\ \hline\hline \end{array} \]

\[\begin{array}{||c|c|c||}\hline\hline
\psi_L=\left(\begin{array}{c}e^-\\\nu_e\\N^0_1\end{array}\right)_L &
\psi_{1L}=\left(\begin{array}{c}
E^-\\ N^0_2\\N^0_3\end{array}\right)_L &
\psi_{2L}=\left(\begin{array}{c} N^0_4\\E^+\\e^+\end{array}\right)_L \\
\hline
(1,3^*,-{1\over 3}) & (1,3^*,-{1\over 3}) & (1,3^*,{2\over
3})\\ \hline\hline\end{array},  \]

\noindent
where the numbers inside the parenthesis refer to
$(SU(3)_c,SU(3)_L,U(1)_X)$ quantum numbers. This anomaly-free structure is
the simplest one we can construct for a single family in $SU(3)_L\otimes
U(1)_X$.  As a matter of fact, the 27 states above are just the 27 states
in the fundamental representation of the electroweak-strong
unification group $E_6$\cite{e6}, so this gauge and fermion structure
is such that $SU(3)_c\otimes SU(3)_L\otimes U(1)_X\subset
E_6$. A phenomenological analysis of this model has been
published already in Ref.\cite{smp}.

\subsubsection{Model B}
For this model we start with an extra up type quark $q=U$ of electric
charge $Q_q=Q_U=2/3 (b=-1/2)$ which in turn implies $Q_l=-1$, that is,
$l_L$ is now an exotic electron $E^-$. Following the same steps as for
model {\bf A} we end up with the following multiplet structure:

\[\begin{array}{||c|c|c|c||}\hline\hline
\chi_L=\left(\begin{array}{c}u\\d\\U \end{array}\right)_L & d^c_L & u^c_L&
U^c_L \\ \hline (3,3,{1\over 3}) & (3^*,1,{1\over 3}) &
(3^*,1,-{2\over 3})
& (3^*,1,-{2\over 3}) \\ \hline\hline \end{array} \]

\[\begin{array}{||c|c|c|c|c|c||}\hline\hline
\psi_L=\left(\begin{array}{c} e^-\\ \nu_e \\E^-_1\end{array}\right)_L
&
\psi_{1L}=\left(\begin{array}{c} N_1^0 \\ E^+_2\\
\nu^c_e\end{array}
\right)_L & \psi_{2L}=\left(\begin{array}{c}
E_2^- \\ N_2^0\\ E^-_3\end{array}\right)_L & e^+_L & E_{1L}^+ &
E_{3L}^+ \\ \hline
(1,3^*,-{2\over 3}) & (1,3^*,{1\over 3}) & (1,3^*,-{2\over 3})
& (1,1,1) & (1,1,1) & (1,1,1) \\ \hline\hline
\end{array}  \]

A simple check shows that this multiplet structure is also free of
anomalies. A phenomenological analysis of this model has been started
already in Ref.\cite{mps}, where it is shown that model {\bf B} as
presented here is a subgroup of $SU(6)\otimes U(1)$, an electroweak-strong
unification group which has not been considered in the literature so far.

The gauge boson content of models {\bf A} and {\bf B} are equivalent, and
they become the same just by replacing $3\leftrightarrow 3^*$ in the
irreducible representations of the fermion fields ($b\rightarrow -b$ when
the complex conjugate of the covariant derivative is taken).

\subsubsection{Other one-family Models}
Following the same steps as for the two previous cases, we attempt to
construct models where $q_L$ has electric charges $-2/3$ or 1/3.
Eq.(\ref{chql}) then implies that $Q_l=1/3$ and $-2/3$ respectively
which correspond to leptons with exotic electric charges.
Not only fractionally charged free particles has not been detected at
low energies, but the phenomenology of those models could become
tremendously confusing with leptons with electric charges equal to
the antiquarks.

In a similar way by asking for a model with $Q_l=1$ we will have,
according to Eq.(\ref{chql}), that $Q_q=-4/3$, a model with a quark with
an exotic electric charge which we have excluded from the models discussed
here (a model with quarks with exotic electric charges is presented in
Ref.\cite{fp} for example).

\subsection{Family models} For these models each individual family
possesses non-vanishing anomalies and the anomaly cancellation takes place
between families and, for some models, only with a matching of the number
of families with the number of quark colors, does the overall anomaly
vanish\cite{fp,long,ozer}. It is also a feature of this type of models
that the third family is treated different to the other two, or either
that all the three families are treated independently.

An algebraic manipulation of Eqs.(\ref{ch}) and (\ref{chql}) and the
anomaly constraint equations, allows us to combine the fermion multiplets
of the two models {\bf A} and {\bf B} to produce the following models
(with the replacement $3\leftrightarrow 3^*$ in model {\bf B} in order to
assure a unique covariant derivative):

\subsubsection{Model C}
All the left-handed lepton generations belong to the representation
$(1,3,-2/3)$ of $(SU(3)_c, SU(3)_L,U(1)_X)$, that is:

\[\begin{array}{||c|c|c||}\hline\hline
\psi^\alpha_L=\left(\begin{array}{c} \nu_\alpha \\ \alpha^- \\E^-_\alpha
\end{array}\right)_L & \alpha^+_L & E^+_{\alpha L}  \\ \hline
(1,3,-2/3) & (1,1,1) & (1,1,1) \\ \hline\hline \end{array} \]

\noindent
for $\alpha = e,\mu,\tau $; while quarks transform as follows:

\[\begin{array}{||c|c|c|c||}\hline\hline
\chi^a_L=\left(\begin{array}{c} d^a\\u^a\\U^a \end{array}\right)_L &
u^{ac}_L & d^{ac}_L& U^{ac}_L \\ \hline
(3,3^*,1/3) & (3^*,1,-{2\over 3}) & (3^*,1,{1\over 3}) &
(3^*, 1,-{2\over 3})  \\ \hline\hline \end{array} \]

\noindent
for $a=1,2$ two of the families. For the other family we have:

\[\begin{array}{||c|c|c|c||}\hline\hline
\chi_{3L}=\left(\begin{array}{c}u_3\\d_3\\D \end{array}\right)_L &
u^c_{3L} & d^c_{3L}& D^c_L \\ \hline
(3, 3,0) & (3^*, 1,-{2\over 3}) & (3^*, 1,{1\over 3}) &
(3^*, 1,{1\over 3})   \\ \hline\hline \end{array}. \]

The arithmetic shows that all the anomalies vanish for this fermion
content. As far as we know the study of this model is
relatively new in the literature; it was introduced for the first time in
Ref.\cite{ozer}.

At first glance this structure does not allow for neutrino masses; even
though, a variant of this model, with the capability to explain the main
features of the atmospheric and solar neutrino experimental
results, has been presented in Ref.\cite{kita1}.

\subsubsection{Model D}
In a similar way we get the following multiplet structure:

\[\begin{array}{||c|c|c|c||}\hline\hline
\chi^a_L=\left(\begin{array}{c} u_a \\ d_{a} \\D_a
\end{array}\right)_L &
u^c_{aL} & d^c_{aL}& D^c_{aL} \\ \hline
(3,3,0) & (3^*, 1,-{2\over 3}) & (3^*, 1,{1\over 3}) &
(3^*, 1,{1\over 3})  \\ \hline\hline \end{array} \]

\noindent
for $a=1,2$, the quarks in two of the three families. For the
quarks in the other family we have:

\[\begin{array}{||c|c|c|c||}\hline\hline
\chi^3_L=\left(\begin{array}{c}d_3\\u_3\\U \end{array}\right)_L &
u^c_{3L} & d^c_{3L}& U^c_{L} \\ \hline
(3, 3^*,{1\over 3}) & (3^*, 1,-{2\over 3}) & (3^*, 1,{1\over 3}) &
(3^*, 1,-{2\over 3})  \\ \hline\hline \end{array}. \]

\noindent The three lepton generations transform now as anti-triplets of
$SU(3)_L$ as follows:

\[\begin{array}{||c|c||}\hline\hline
\psi^\alpha_L=\left(\begin{array}{c} \alpha^-\\ \nu_\alpha \\ N^0_\alpha
\end{array}\right)_L &
\alpha^+_L  \\ \hline
(1, 3^* , -{1\over 3}) & (1,1,1) \\ \hline\hline \end{array}, \]

\noindent
for $\alpha = e, \mu ,\tau$ the three families. This model has been
largely studied in the literature (see Refs.\cite{long}). Again, this
model has been used recently in connection with neutrino
oscillations\cite{kita2}.

\subsubsection{Other models}
Contrary to the one family models, we can now play the game of canceling
the anomalies in several different ways.

We start by defining the following closed set of fermions (closed in the
sense that they include the antiparticles of the charged particles):\\
$S_1=[(\nu_\alpha, \alpha^- ,E_\alpha^-); \alpha^+;
E_\alpha^+]$ with
quantum numbers $(1,3,-2/3); (1,1,1)$ and $(1,1,1)$ respectively.\\
$S_2=[(\alpha^- , \nu_\alpha ,N_\alpha^0); \alpha^+]$ with
quantum numbers $(1,3^*,-1/3)$ and $(1,1,1)$ respectively.\\
$S_3=[(d,u,U);u^c;d^c;U^c]$ with
quantum numbers $(3,3^*,1/3); (3^*,1,-2/3);(3^*,1,1/3)$ and
$(3^*,1,-2/3)$ respectively.\\
$S_4=[(u,d,D); d^c;u^c;D^c]$ with
quantum numbers $(3,3,0); (3^*,1,1/3); (3^*,1,-2/3)$ and
$(3^*,1,1/3)$ respectively.\\
$S_5=[(e^-,\nu_e ,N_1^0); (E^-,N_2^0,N_3^0)$; $(N_4^0,E^+,e^+)]$ with
quantum numbers $(1,3^*,-1/3)$; $(1,3^*,-1/3)$ and
$(1,3^*,2/3)$ respectively.\\
$S_6=[(\nu_e, e^-, E^-); (E_2^+, N_1^0,N_2^0);(N_3^0, E_2^- ,E_3^-);
e^+,
E_1^+;E_3^+]$ with quantum numbers
$(1,3,-2/3)$; $(1,3,1/3)$; $(1,3,-2/3);(1,1,1);(1,1,1)$ and
$(1,1,1)$ respectively.

Now we calculate the four anomalies for each set of particles. The
results are presented in Table I.

\begin{center}

TABLE I. Anomalies for $S_i$.

\begin{tabular}{||l||c|c|c|c|c|c||}\hline\hline
Anomalies           & $S_1$& $S_2$& $S_3$& $S_4$& $S_5$ & $S_6$ \\
\hline\hline
$[SU(3)_c]^2U(1)_X$ & 0    &  0    &  0    &  0   &  0  & 0 \\
$[SU(3)_L]^2U(1)_X$ & $-2/3$ & $ -1/3$ &  1    &  0   &  0  & $-1$ \\
$[grav]^2U(1)_X$     & 0    &  0    &  0    &  0   &  0  & 0 \\
$[U(1)_X]^3$        & 10/9 & 8/9   & $-12/9$ & $-6/9$ & 6/9 & 12/9 \\
\hline\hline
\end{tabular}
\end{center}

\vspace{.5cm}


Notice from Table I that model {\bf A} is just $(S_4+S_5)$ and model
{\bf B} is $(S_3+S_6)$. Model {\bf C} is represented by $(3S_1+2S_3+S_4)$
and model {\bf D} by $(3S_2+S_3+2S_4)$, but what is most remarkable is
that we can now construct new anomaly-free models for two, three, four
and more families. For example two new three family models are:\\
{\bf Model E}: $S_1+S_2+S_3+S_4$ plus Model {\bf A} =
$(S_1+S_2+S_3+2S_4+S_5)$\\
{\bf Model F}  $S_1+S_2+S_3+S_4$ plus Model {\bf
B}=$(S_1+S_2+2S_3+S_4+S_6)$.\\
A model for four families will be given for example by:
$2(S_1+S_2+S_3+S_4)$, etc..

The main feature of models {\bf E} and {\bf F} above is that, contrary to
models {\bf C} and {\bf D}, each one of the three families is treated in a
different way. As far as we know, these two models have not been studied
in the literature so far.

\section{The scalar sector}
Even though the representation content for the fermions may vary
significantly from model to model, all $SU(3)_L\otimes U(1)_X$ models
presented so far have the same gauge boson sector as it will be discussed
in the following section. Now, to achieve an spontaneous breaking of the
symmetry in the most economic way, using the chain

\[SU(3)_c\otimes SU(3)_L\otimes U(1)_X\longrightarrow
SU(3)_c\otimes SU(2)_L\otimes U(1)_Y\longrightarrow SU(3)_c\otimes
U(1)_Q,\]

\noindent
we need two complex higgs scalars
$\phi_i(1,3^*,-1/3)=(\phi_i^-,\phi_i^0, \phi_i^{'0}),\; i=1,2$,
with Vacuum Expectation Values (VEV)
$\langle\phi_1\rangle=(0,0,V)^T$ and
$\langle\phi_2\rangle=(0,v/\sqrt{2},0)^T$, with the hierarchy $V>>v\sim
250$ GeV the electroweak mass scale. Now, to break the symmetry and
at the same time give masses to all the fermion fields is a model
dependent analysis. So, let us outline in the following three sections the
analysis for model {\bf A} for example, for which a third higgs field
$\phi_3(1,3^*,2/3)=(\phi_3^0,\phi_3^+,\phi_3^{'+})$ with VEV
$\langle\phi_3\rangle=(v'/\sqrt{2},0,0)^T$ is
also needed, where $v'\simeq v$\cite{smp}.

The analysis for model {\bf B} is done in Ref.\cite{mps} and for model
{\bf D} in Ref.\cite{long}.

\section{The gauge boson sector}
There are a total of 17 gauge bosons in the gauge group under
consideration; they are: one gauge
field $B^\mu$ associated with $U(1)_X$, the 8 gluon fields
associated with $SU(3)_c$ which remain massless after breaking the
symmetry, and another 8 associated
with $SU(3)_L$ and that we write for convenience in the following way:
\[{1\over 2}\lambda_\alpha A^\mu_\alpha={1\over \sqrt{2}}\left(
\begin{array}{ccc}D^\mu_1 & W^{+\mu} & K^{+\mu} \\ W^{-\mu} &
D^\mu_2 &  K^{0\mu} \\
K^{-\mu} & \bar{K}^{0\mu} & D^\mu_3 \end{array}\right), \]
where $D^\mu_1=A_3^\mu/\sqrt{2}+A_8^\mu/\sqrt{6},\;
D^\mu_2=-A_3^\mu/\sqrt{2}+A_8^\mu/\sqrt{6}$,
and $D^\mu_3=-2A_8^\mu/\sqrt{6}$. $\lambda_i, \; i=1,2,...,8$ are the eight
Gell-Mann matrices normalized as mentioned in Section {\bf 2.1} This allows
us to write now the charge operator as
\[Q=\frac{\lambda_3}{2}+\frac{\lambda_8}{2\sqrt{3}}+XI_3,\]
where $I_3$ is the $3\times 3$ unit matrix.

In model {\bf A}, after breaking the symmetry with
$\langle\phi_i\rangle ,\; i=1,2,3$, and
using for the covariant derivative for
triplets $D^\mu=\partial^\mu-i{g\over 2}
\lambda_\alpha A^\mu_\alpha-ig'XB^\mu$, we get the following mass terms for
the charged gauge bosons in the electroweak sector:
$M^2_{W^\pm}={g^2\over 4}(v^2+v'^2), \; M^2_{K^\pm}={g^2\over 4}(2V^2+v'^2), \;
M^2_{K^0(\bar{K}^0)}={g^2\over 4}(2V^2+v^2)$. For the neutral gauge bosons
we get a mass term of the form:
\[M=V^2(\frac{g'B^\mu}{3}-\frac{gA_8^\mu}{\sqrt{3}})^2
+ \frac{v^2}{8}(\frac{2g'B^\mu}{3}-gA^\mu_3 +\frac{gA_8^\mu}{\sqrt{3}})^2
+\frac{v'^2}{8}(gA_3^\mu-\frac{4g'B^\mu}{3}+\frac{gA^\mu_8}{\sqrt{3}})^2\]

By diagonalizing $M$ we get the physical neutral gauge bosons which are
defined through the mixing angle $\theta$ and $Z_\mu,\; Z'_\mu$ by:
\begin{eqnarray}\nonumber
Z_1^\mu&=&Z_\mu \cos\theta+Z'_\mu \sin\theta \\ \nonumber
Z_2^\mu&=&-Z_\mu \sin\theta+Z'_\mu \cos\theta \\ \nonumber
-\tan(2\theta)&=&\frac{\sqrt{12}C_W(1-T_W^2/3)^{1/2}[v'^2(1+T_W^2)-v^2(1-T_W^2)]}
{3(1-T_W^2/3)(v^2+v'^2)-C_W^2[8V^2+v^2(1-T_W^2)^2+v'^2(1+T_W^2)^2]},\\ \label{tant}
\end{eqnarray}
where the photon field $A^\mu$ and the fields $Z_\mu$ and $Z'_\mu$ are given by
\begin{eqnarray} \nonumber
A^\mu&=&S_W A_3^\mu + C_W\left[\frac{T_W}{\sqrt{3}}A_8^\mu+
(1-T_W^2/3)^{1/2}B^\mu\right],\\ \nonumber
Z^\mu&=& C_W A_3^\mu - S_W\left[\frac{T_W}{\sqrt{3}}A_8^\mu+
(1-T_W^2/3)^{1/2}B^\mu\right],\\ \label{foton}
Z'^\mu&=&-(1-T_W^2/3)^{1/2}A_8^\mu+\frac{T_W}{\sqrt{3}}B^\mu.
\end{eqnarray}
$S_W$ and $C_W$ are the sine and cosine of the electroweak mixing
angle respectively ($T_W=S_W/C_W$) defined by
$S_W=\sqrt{3}g'/\sqrt{3g^2+4g'^2}$. Also we can identify the $Y$
hypercharge associated with the SM gauge boson as:
\[Y^\mu=\left[\frac{T_W}{\sqrt{3}}A_8^\mu+ (1-T_W^2/3)^{1/2}B^\mu\right].\]
In the limit $\theta\longrightarrow 0$, $M_{Z}=M_{W^{\pm}}/C_W$, and
$Z_1^\mu=Z^\mu$ is the gauge boson of the SM. This limit
is obtained either by demanding $V\longrightarrow\infty$ or
$v'^2=v^2(C_W^2-S_W^2)$. In general $\theta$ may be
different from zero although it takes a very small value, determined from
phenomenology for each particular model.

The former definitions for $A^\mu,\, Z^\mu,\, Z^{'\mu},\, Y^\mu$ and $S_W$
are the same for all the six models in Section {\bf 2}. The value for
$\tan (2\theta)$ and the expressions for the masses of the gauge bosons
are model dependent.

\subsection{Currents}
The  currents for fermions are different for each model and also
they are different from those of the SM. As an example let us present the
analysis for model {\bf A}\cite{smp}; a similar analysis for model
{\bf B} is presented in Ref.\cite{mps} and for model {\bf D}  in
Ref.\cite{long}.

\subsubsection{Charged currents}
The interactions among the charged vector fields with leptons for model
{\bf A} are
\begin{eqnarray}\nonumber
H^{CC}&=&{g\over \sqrt{2}}[W^+_\mu(\bar{u}_L\gamma^\mu d_L-
\bar{\nu}_{eL}\gamma^\mu e^-_L-\bar{N}^0_{2L}\gamma^\mu E^-_L-
\bar{E}^+_L\gamma^\mu N^0_{4L}) \\ \nonumber
& & +K^+_\mu(\bar{u}_L\gamma^\mu D_L-
\bar{N}^0_{1L}\gamma^\mu e^-_L-\bar{N}^0_{3L}\gamma^\mu E^-_L-
\bar{e}^+_L\gamma^\mu N^0_{4L}) \\
& & +K^0_\mu(\bar{d}_L\gamma^\mu D_L-
\bar{N}^0_{1L}\gamma^\mu \nu_{eL}-\bar{N}^0_{3L}\gamma^\mu N^0_{2L}-
\bar{e}^+_L\gamma^\mu E^+_L)] + H.c.,
\end{eqnarray}
which implies that the interactions with $K^\pm$ and
$K^0(\bar{K}^0)$ bosons violate the lepton number and the weak isospin.
Notice also that the first two terms in the previous expression constitute
the charged weak current of the SM as far as we identify $W^\pm$ as the
$SU(2)_L$ charged left-handed weak bosons.

\subsubsection{Neutral currents}
The neutral currents $J_\mu(EM),\; J_\mu(Z)$ and $J_\mu(Z')$,  associated
with the Hamiltonian $H^0=eA^\mu J_\mu(EM)+{g\over {C_W}}Z^\mu J_\mu(Z) +
{g'\over \sqrt{3}}Z'^\mu J_\mu(Z')$ are:
\begin{eqnarray}\nonumber
J_\mu(EM)&=&{2\over 3}\bar{u}\gamma_\mu u-{1\over 3}\bar{d}\gamma_\mu d
-{1\over 3}\bar{D}\gamma_\mu D- \bar{e}^-\gamma_\mu e^--
\bar{E}^-\gamma_\mu E^-=\sum_f q_f\bar{f}\gamma_\mu f \\ \nonumber
J_\mu(Z)&=&J_{\mu,L}(Z)-S^2_WJ_\mu(EM)\\
J_\mu(Z')&=&T_WJ_\mu(EM)-J_{\mu,L}(Z'),
\end{eqnarray}
where $e=gS_W=g'C_W\sqrt{1-T_W ^2/3}>0$ is the electric charge, $q_f$
is the electric charge of the fermion $f$ in units of $e$, $J_\mu(EM)$
is the electromagnetic current (vectorlike as it should be), and the
left-handed currents are
\begin{eqnarray} \nonumber
J_{\mu,L}(Z)&=&{1\over 2}(\bar{u}_L\gamma_\mu u_L-
\bar{d}_L\gamma_\mu d_L+\bar{\nu}_{eL}\gamma_\mu \nu_{eL}-
\bar{e}^-_L\gamma_\mu e^-_L+\bar{N}^0_{2}\gamma_\mu N^0_{2}
-\bar{E}^-\gamma_\mu E^-) \\ \nonumber
&=&\sum_f T_{3f}\bar{f}_L\gamma_\mu f_L \\ \nonumber
J_{\mu,L}(Z')&=&S_{2W}^{-1}(\bar{u}_L\gamma_\mu u_L -
\bar{e}^-_L\gamma_\mu e^-_L-\bar{E}^-_L\gamma_\mu E^-_L-
\bar{N}^0_{4L}\gamma_\mu N^0_{4L})\\ \nonumber
& & T_{2W}^{-1}(\bar{d}_L\gamma_\mu d_L-\bar{E}^+_L\gamma_\mu E^+_L -
\bar{\nu}_{eL}\gamma_\mu \nu_{eL} -\bar{N}^0_{2L}\gamma_\mu N^0_{2L}) \\ \nonumber
& &- T_W^{-1}(\bar{D}_L\gamma_\mu D_L - \bar{e}^+_L\gamma_\mu e^+_L
- \bar{N}^0_{1L}\gamma_\mu N^0_{1L}-\bar{N}^0_{3L}\gamma_\mu N^0_{3L}) \\
&=&\sum_f T_{9f}\bar{f}_L\gamma_\mu f_L,
\end{eqnarray}
where $S_{2W}=2S_WC_W,\; T_{2W}=S_{2W}/C_{2W}, \; C_{2W}=C_W^2-S_W^2, \;
\bar{N}^0_{2}\gamma_\mu N^0_{2}=\bar{N}^0_{2L}\gamma_\mu N^0_{2L}
+\bar{N}^0_{2R}\gamma_\mu N^0_{2R}= \bar{N}^0_{2L}\gamma_\mu N^0_{2L}
-\bar{N}^{0c}_{2L}\gamma_\mu N^{0c}_{2L}= \bar{N}^0_{2L}\gamma_\mu N^0_{2L}
-\bar{N}^0_{4L}\gamma_\mu N^0_{4L}$, and similarly
$\bar{E}\gamma_\mu E=\bar{E}^-_L\gamma_\mu E^-_L -
\bar{E}^+_L\gamma_\mu E^+_L$.
In this way $T_{3f}=Dg.(1/2,-1/2,0)$ is the
third component of the weak isospin acting on the representation 3 of
$SU(3)_L$ (the negative when acting on $3^*$), and
$T_{9f}=Dg.(S_{2W}^{-1}, T_{2W}^{-1}, -T_W^{-1})$ is a convenient
$3\times 3$ diagonal matrix acting on the representation 3 of $SU(3)_L$
(the negative when acting on $3^*$).
Notice that $J_\mu(Z)$ is just the generalization of the neutral current
present in the SM, which allows us to identify $Z_\mu$ as the
neutral gauge boson of the SM.

The couplings of the physical states $Z_1^\mu$ and $Z_2^\mu$ are then
given by:
\begin{eqnarray} \nonumber
H^{NC}&=&\frac{g}{2C_W}\sum_{i=1}^2Z_i^\mu\sum_f\{\bar{f}\gamma_\mu
[a_{iL}(f)(1-\gamma_5)+a_{iR}(f)(1+\gamma_5)]f\} \\
      &=&\frac{g}{2C_W}\sum_{i=1}^2Z_i^\mu\sum_f\{\bar{f}\gamma_\mu
      [g(f)_{iV}-g(f)_{iA}\gamma_5]f\},
\end{eqnarray}
where
\begin{eqnarray}  \nonumber
a_{1L}(f)&=&\cos\theta(T_{3f}-q_fS^2_W)-\frac{g'\sin\theta C_W}{g\sqrt{3}}
(T_{9f}-q_fT_W) \\ \nonumber
a_{1R}(f)&=&-q_fS_W(\cos\theta S_W-\frac{g'\sin\theta}{g\sqrt{3}})\\ \nonumber
a_{2L}(f)&=&-\sin\theta(T_{3f}-q_fS^2_W)-\frac{g'\cos\theta C_W}{g\sqrt{3}}
(T_{9f}-q_fT_W) \\ \label{coupa}
a_{2R}(f)&=&q_fS_W(\sin\theta S_W+\frac{g'\cos\theta}{g\sqrt{3}}),
\end{eqnarray}
and
\begin{eqnarray} \nonumber
g(f)_{1V}&=&\cos\theta(T_{3f}-2S_W^2q_f)-\frac{g'\sin\theta}{g\sqrt{3}}
(T_{9f}C_W-2q_fS_W) \\ \nonumber
g(f)_{2V}&=&-\sin\theta(T_{3f}-2S_W^2q_f)-\frac{g'\cos\theta}{g\sqrt{3}}
(T_{9f}C_W-2q_fS_W) \\ \nonumber
g(f)_{1A}&=&\cos\theta T_{3f}-\frac{g'\sin\theta}{g\sqrt{3}}T_{9f}C_W \\ \label{coupg}
g(f)_{2A}&=&-\sin\theta T_{3f}-\frac{g'\cos\theta}{g\sqrt{3}}T_{9f}C_W,
\end{eqnarray}
to be compared with the SM values $g(f)_{1V}^{SM}=T_{3f}-2S_Wq_f$ and
$g(f)_{1A}^{SM}=T_{3f}$.
The values of $g_{iV},\; g_{iA};\;\; i=1,2$ are listed in Tables II and
III. As we can see, in the limit $\theta=0$ the couplings of $Z_1^\mu$ to
the  ordinary leptons and quarks are the same as in the SM. Because of
this, we can test the new phenomenology beyond the SM.

\pagebreak

\begin{center}
TABLE II. The $Z_1^\mu\longrightarrow \bar{f}f$ couplings.
\begin{tabular}{||l||c|c||}\hline\hline
f& $g_{1V}$ & $g_{1A}$ \\ \hline\hline
u& $({1\over 2}-{4S_W^2\over 3})[\cos\theta-\sin\theta
/(4C_W^2-1)^{1/2}]$ &
${\cos\theta\over 2}-\sin\theta/[2(4C_w^2-1)^{1/2}]$ \\ \hline
d& $\cos\theta (-{1\over 2}+{2S_W^2 \over 3})-
\frac{\sin\theta}{(4C_W^2-1)^{1/2}}({1\over 2} - {S_W^2\over 3})$ &
$-{1\over 2}\{\cos\theta + \sin\theta C_{2W}/[2(4C_W^2-1)^{1/2}]\}$ \\ \hline
D& ${2S_W^2\cos\theta \over 3}+\sin\theta (1-{5\over 3}S_W^2)/(4C_W^2-1)^{1/2}$ &
$C_W^2\sin\theta /(4C_W^2-1)^{1/2}$ \\ \hline
$e^-$& $\cos\theta (-{1\over 2}+2S_W^2)+
\frac{3\sin\theta}{(4C_W^2-1)^{1/2}}({1\over 2} - S_W^2)$ &
$ -{\cos\theta\over 2}+\frac{\sin\theta}{(4C_W^2-1)^{1/2}}({1\over 2}-C_W^2)$
 \\ \hline
$E^-$& $\cos\theta (-1+2S_W^2)-
\frac{S_W^2\sin\theta}{(4C_W^2-1)^{1/2}}$ &
$C_W^2\sin\theta /(4C_W^2-1)^{1/2}$ \\ \hline
$\nu_e,\; N_2^0$ &
${1\over 2}[\cos\theta+\sin\theta(1-2S_W^2)/(4C_W^2-1)^{1/2}] $ &
${1\over 2}(\cos\theta+\sin\theta(1-2S_W^2)/(4C_W^2-1)^{1/2} $ \\ \hline
$N_1^0,\; N_3^0$ &
$-C_W^2\sin\theta /(4C_W^2-1)^{1/2}$& $-C_W^2\sin\theta /(4C_W^2-1)^{1/2}$
\\ \hline
$N_4^0$ & $-{1\over 2}[\cos\theta-\sin\theta/(4C_W^2-1)^{1/2}]$ &
$-{1\over 2}[\cos\theta-\sin\theta/(4C_W^2-1)^{1/2}]$ \\ \hline\hline
\end{tabular}
\end{center}

\vspace{1cm}

\begin{center}
TABLE III. The $Z_2^\mu\longrightarrow \bar{f}f$ couplings.
\begin{tabular}{||l||c|c||}\hline\hline
f& $g_{2V}$ & $g_{2A}$ \\ \hline\hline
u& $({1\over 2}-{4S_W^2\over 3})[-\sin\theta-\cos\theta /(4C_W^2-1)^{1/2}]$&
${-\sin\theta\over 2}-\cos\theta/[2(4C_w^2-1)^{1/2}]$ \\ \hline
d& $-\sin\theta (-{1\over 2}+{2S_W^2 \over 3})-
\frac{\cos\theta}{(4C_W^2-1)^{1/2}}({1\over 2} - {S_W^2\over 3})$ &
$-{1\over 2}\{-\sin\theta + \cos\theta C_{2W}/[2(4C_W^2-1)^{1/2}]\}$ \\ \hline
D& ${-2S_W^2\sin\theta \over 3}+\cos\theta (1-{5\over 3}S_W^2)/
(4C_W^2-1)^{1/2}$ &
$C_W^2\cos\theta /(4C_W^2-1)^{1/2}$ \\ \hline
$e^-$& $-\sin\theta (-{1\over 2}+2S_W^2)+
\frac{3\cos\theta}{(4C_W^2-1)^{1/2}}({1\over 2} - S_W^2)$ &
$ {\sin\theta\over 2}+\frac{\cos\theta}{(4C_W^2-1)^{1/2}}({1\over 2}-C_W^2)$
 \\ \hline
$E^-$& $-\sin\theta (-1+2S_W^2)-
\frac{S_W^2\cos\theta}{(4C_W^2-1)^{1/2}}$ &
$C_W^2\cos\theta /(4C_W^2-1)^{1/2}$ \\ \hline
$\nu_e,\; N_2^0$ &
${1\over 2}[-\sin\theta+\cos\theta(1-2S_W^2)/(4C_W^2-1)^{1/2}] $ &
${1\over 2}(-\sin\theta+\cos\theta(1-2S_W^2)/(4C_W^2-1)^{1/2} $ \\ \hline
$N_1^0,\; N_3^0$ &
$-C_W^2\cos\theta /(4C_W^2-1)^{1/2}$& $-C_W^2\cos\theta /(4C_W^2-1)^{1/2}$
\\ \hline
$N_4^0$ & ${1\over 2}[\sin\theta+\cos\theta/(4C_W^2-1)^{1/2}]$ &
${1\over 2}[\sin\theta+\cos\theta/(4C_W^2-1)^{1/2}]$ \\ \hline\hline
\end{tabular}
\end{center}


\section{Masses for fermions}
Again this subject is model dependent. Just for the sake of completeness
let us write the Yukawa lagrangian that the three higgs scalars in Section
{\bf 3} produce for the fermion fields in model {\bf A}\cite{smp}:
\begin{eqnarray} \nonumber
{\cal L}_Y&=&{\cal L}^Q_Y+{\cal L}^l_Y\\
{\cal L}_Y^Q&=&\chi^T_LC(h_u\phi_3u_L^c+h_D\phi_1D_L^c+h_d\phi_2d_L^c+
               h_{dD}\phi_2D_L^c+h_{Dd}\phi_1d_L^c) + h.c.\\ \nonumber
{\cal L}^l_Y&=&\epsilon_{abc}[\psi_L^aC(h_1\psi_{1L}^b\phi_3^c
             +h_2\psi_{2L}^b\phi_1^c+h_3\psi_{2L}^b\phi_2^c) +
            \psi_{1L}^aC(h_4\psi_{2L}^b\phi_1^c+h_5\psi_{2L}^b\phi_2^c)]\\
            & & + H.c.,
\end{eqnarray}
where $h_\eta , \; \eta=u,d,D,dD,Dd,1,2,3,4,5$ are Yukawa couplings of
order one, $a,b,c$ are $SU(3)_L$ tensor indices and $C$ is the charge
conjugation operator.

Using the VEV as in section 3 and assuming that we are referring to the
third family, we see that $m_t=h_uv'/\sqrt{2}$, $m_D\sim h_DV$ but it
mixes with the $b$ quark producing a kind of see-saw
mechanism\cite{seesaw} that implies $m_b<<m_t$. Also for leptons we have
$m_E\sim h_4V$ but again it mixes with the $\tau$ lepton producing also a
kind of see saw mechanism which implies that $m_\tau\sim m_b<<m_t$. The
neutral sector is more complicated; the analysis of the $5\times 5$ mass
matrix gives: first two eigenvalues $\pm h_1v'/\sqrt{2}$ which correspond
to a Dirac neutrino with a mass of the order of the electroweak mass
scale; other two are $\pm V +\eta$,
where $\eta$ is a small see-saw quotient, which correspond to a
very massive pseudo-Dirac neutrino, and finally a tiny mass Majorana
neutrino.

So the higgs fields and VEV used break the symmetry in the appropriate
way, and produce a realistic pattern of masses for the fermion fields (at
least for one of the families).

\section{Conclusions}
In this paper we have studied the theory of
$SU(3)c\otimes SU(3)_L\otimes U(1)_X$ in detail. By restricting the
fermion field representations to particles without exotic electric charges
we end up with six different models, two one family models and four models
for three families. The two one family models are sketched in the papers
by K.T. Mahanthappa and P.K. Mohapatra in Ref.\cite{fallet}, but enough
attention was not paid to the anomaly cancellation constraints in their
analysis. The four three family models are relatively new in the
literature, with two of them (models {\bf E} and {\bf F}) introduced
here for the first time, as far as we know.

If we allow for particles with exotic electric charges in our analysis, we
end up with an infinite number of models, where the model in
Refs.\cite{fp} is just one of them (probably the most elegant one!).

The low energy predictions of the six models discussed here are not the
same. All of them have in common a new neutral current which mixes with
the SM neutral current which is also included as part of each model. When
the mixing angle between the two neutral currents is zero
($\sin\theta=0)$, exact agreement with the SM predictions is achieved, but
the use of experimental results from LEP, SLAC and atomic parity violation
bound the mixing angle to values which are model dependent. For partial
analysis see for example Refs.\cite{long, smp} and \cite{mps}.

Detailed analysis in each model of flavor changing neutral currents, GIM
mechanism, mass scales of the new gauge bosons, mass spectrum for the
neutral spin 1/2 particles etc., are model dependent and they will be
presented elsewhere.

Finally let us mention that the most remarkable result of our analysis is
the existence of models {\bf E} and {\bf F}, where the three families are
treated different. In these models it should be simple to implement the
horizontal survival hypothesis\cite{hsh}, that is, to provide masses at
tree level only for the particles in the third family, as done for example
in the previous section, with the known particles in the first and second
families getting masses as radiative corrections.

\section{Acknowledgments}
This work was partially supported by BID and Colciencias in Colombia. We
thank C. Garc\'\i a Canal for a critical reading of the original manuscript. WAP acknowledges warm hospitality from the Theoretical Physics Laboratory at the Universidad de la Plata in Argentina, where this work was
completed.

\end{document}